\documentclass[aps, pra, amsmath, amsfonts, amssymb, showpacs, floatfix, preprintnumbers, superscriptaddress, twocolumn, 10pt, a4paper]{revtex4-1}
\usepackage{graphicx}
\usepackage{epstopdf}
\usepackage[utf8]{inputenc}
\usepackage[T1]{fontenc}
\usepackage{lmodern}
\usepackage[]{xcolor}
\usepackage[normalem]{ulem}

\usepackage[breaklinks 					
,colorlinks 							
,citecolor=blue 						
,linkcolor=blue 						
,urlcolor=blue 							
,bookmarks 								
,bookmarksnumbered						
,hyperfootnotes=false 					
,pdfpagelabels
,pdfstartview=FitH 						
,pdfpagemode=UseOutlines				
,bookmarksopenlevel=2 					
]{hyperref}
 	
\newcommand{\kfa}{$\left(k_\mathrm F a\right)^{-1}$}

\begin{document}
	
\title{Pair fraction in a finite temperature Fermi gas on the BEC side of the BCS-BEC crossover}

\author{Thomas Paintner}
\author{Daniel K. Hoffmann}
\author{Manuel J\"{a}ger}
\author{Wolfgang Limmer}
\author{Wladimir Schoch}
\author{Benjamin Deissler}
\affiliation{Institut f\"{u}r Quantenmaterie and Center for Integrated Quantum Science and Technology ($\mathrm{IQ}^{ST}$), Universit\"{a}t Ulm, 89069 Ulm, Germany}
\author{Michele Pini}
\affiliation{School of Science and Technology, Physics Division, Universit\`{a} di Camerino, 62032 Camerino (MC), Italy}
\author{Pierbiagio Pieri}
\affiliation{School of Science and Technology, Physics Division, Universit\`{a} di Camerino, 62032 Camerino (MC), Italy}
\affiliation{INFN, Sezione di Perugia, 06123 Perugia (PG), Italy}
\author{Giancarlo Calvanese Strinati}
\affiliation{School of Science and Technology, Physics Division, Universit\`{a} di Camerino, 62032 Camerino (MC), Italy}
\affiliation{INFN, Sezione di Perugia, 06123 Perugia (PG), Italy}
\affiliation{CNR-INO, Istituto Nazionale di Ottica, Sede di Firenze, 50125 Firenze (FI), Italy}
\author{Cheng Chin}
\affiliation{James Franck Institute and Enrico Fermi Insitute, University of Chicago, 60637 Chicago, USA}
\author{Johannes Hecker Denschlag}
\affiliation{Institut f\"{u}r Quantenmaterie and Center for Integrated Quantum Science and Technology ($\mathrm{IQ}^{ST}$), Universit\"{a}t Ulm, 89069 Ulm, Germany}

\date{\today}

\begin{abstract}
We investigate pairing in a strongly interacting two-component Fermi gas with positive scattering length. In this regime, pairing occurs at temperatures above the superfluid critical temperature; unbound fermions and pairs coexist in thermal equilibrium. Measuring the total number of these fermion pairs in the gas we systematically investigate the phases in the sectors of pseudogap and preformed-pair. Our measurements quantitatively test predictions from two theoretical models. Interestingly, we find that already a model based on classical atom-molecule equilibrium describes our data quite well.
\end{abstract}

\pacs{03.75.Ss, 67.10.Db, 67.85.Lm}

\maketitle
\section{Introduction}
A unique feature of fermionic superfluids is the pairing. For a weakly interacting Bardeen-Cooper-Schrieffer (BCS) superfluid pairing occurs directly at the critical temperature for superfluidity $T_c$ \cite{BCS}. This pairing is accompanied with the emergence of an excitation gap $\Delta_{sc}$ which is identified with the superfluid order parameter and $\Delta_{sc}^2$ is proportional to the density of condensed pairs \cite{BCS2}. For fermions with strong coupling, an excitation gap already emerges at a temperature above $T_c$. This is referred to as the pseudogap regime \cite{levin_gap_review}. The existence of the pseudogap has been observed early on, e.g. in underdoped high-$T_c$ superconductors \cite{phasediagram-hightc,pseudogap_hightc}. While its nature has been intensely studied, it is still not fully understood. Understanding the pseudogap is expected to be the key for revealing the mechanism  behind high-$T_c$ superconductivity \cite{pseudogap_dissc,mueller_pgap}. One interpretation of the pseudogap is based on the presence of non-condensed pairs with non-vanishing momentum \cite{levin-hulet_mol}.

Ultracold Fermi gases are an excellent system for investigating the gap and pseudogap physics from the BCS to Bose-Einstein condensate (BEC) regimes \cite{levin_pseudogap}. Using radio-frequency (RF) spectroscopy in various forms, e.g. \cite{Chin1128,ketterle_gap,gaebler-2010,debyjin_pseudogap}, the excitation gap  has been studied in the way similar to angle-resolved photoemission spectroscopy (ARPES) of solid state systems \cite{gap_cuprates}. Evidence for pairing above $T_c$ was found in the RF experiments, as well as in other physical quantities, such as viscosity \cite{thomas_viscosity}, heat capacity \cite{thomas_heatcapacity}, and  Tan's contact \cite{strinati_contact,vale_contact}.

In this article, we investigate pairing of fermions for various temperatures and interaction strengths on the BEC side of the BEC-BCS crossover. For this, we measure the total number of bound fermion pairs $N_p$ in our sample for $T > T_c$. Such counting of fermion pairs is in general not possible for solid state systems and therefore complements existing methods. We determine the fermion pair number by converting all atom pairs to tightly-bound diatomic molecules, either by  photoexcitation \cite{partridge2005molecular} or by a fast magnetic-field ramp \cite{Regal2004,Zwierlein2004} and  measuring the decrease in atom number of the cloud. When we compare the measured and calculated pair numbers we find quite good agreement with two models: an {\em ab initio} $t$-matrix approach and a classical statistical model of atom-molecule equilibrium \cite{chin}. We provide an explanation why the classical model achieves good results, despite the fact that strong interactions and quantum statistics play an important role in our system.

\begin{figure}[t]
	\includegraphics[width=\columnwidth]{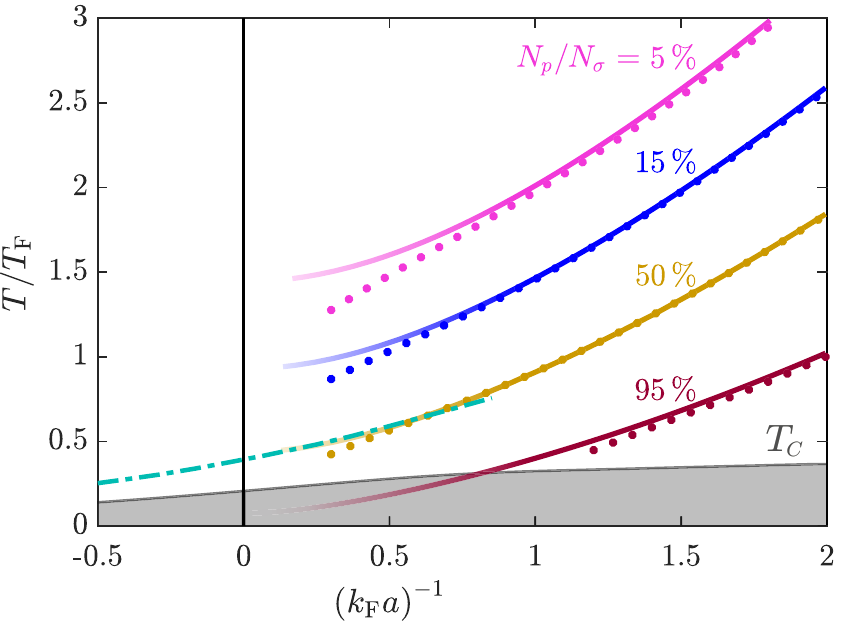}	
	\caption{ Theoretical phase diagram for a balanced two-component harmonically trapped ultracold Fermi gas in the vicinity of a Feshbach resonance (vertical line) where $k_F$ and $T_F$ are determined in the trap center. Shown are calculated contours for various pair fractions. Dotted lines are based on a self-consistent $t$-matrix approach \cite{theoretical_paper}, while solid lines are based on a classical model of non-interacting atoms and molecules (see text) \cite{chin}. Close to the Feshbach resonance the solid lines are blurred because the classical model is expected to lose its validity. The cyan dash-dotted line marks a pair breaking temperature, as calculated by \cite{strinati_Tc} with a BCS mean field model that was extended to the near-BEC regime. The gray shaded area marks the superfluid phase below the critical temperature $T_c$ that was calculated within the self-consistent $t$-matrix approach \cite{theoretical_tcpaper}.}
	\label{fig:skizze}
\end{figure}

In the following, we consider an ultracold, spin-balanced, strongly-interacting two-component Fermi gas in a harmonic trap. Collisions lead to pairing of atoms with opposite spins, $\left| \uparrow \right\rangle$, $\left| \downarrow \right\rangle$. For a given temperature and interaction strength well defined fractions of pairs and atoms are established at thermal equilibrium, as long as collisional losses are negligible. Figure \ref{fig:skizze} shows the phase diagram of such a system in the vicinity of a Feshbach resonance at  $\left(k_\mathrm F a\right)^{-1} = 0$. Here, $a$ is the s-wave scattering length, $k_\mathrm F = \sqrt{2m E_\mathrm F}/\hbar$ denotes the norm of the Fermi wave vector, $m$ is the atomic mass, and $E_\mathrm F=k_\mathrm B T_\mathrm F$ is the Fermi energy in the trap center with $k_\mathrm B$ the Boltzmann constant. The dash-dotted and solid lines are contours of constant molecular fractions $N_p/N_\sigma$ for two different approaches. Here, $N_\sigma = N_p + N_a$ is the number of all atoms per spin state regardless whether they are bound in pairs ($N_p$) or free ($N_a$). The dotted lines are calculations based on a self-consistent $t$-matrix approach \cite{theoretical_paper}, while the solid lines correspond to a statistical mechanics approach treating the particles as  a canonical ensemble of non-interacting molecules and atoms in chemical equilibrium (see \cite{chin} and Appendix \ref{app:model}). Here, the molecules have a binding energy of $E_b = -\hbar^2/(m a^2)$. Also shown is a calculation (cyan dash-dotted line) by  Perali \textit{et al.} \cite{strinati_Tc} of the BCS mean-field critical temperature which provides an approximate estimate of the pair breaking temperature. It partially coincides with the $50\,\%$ pair fraction line of the statistical mechanics approach.

We carry out our experiments with a spin-balanced two-component Fermi gas of ${}^6 \mathrm{Li}$ atoms which is initially prepared at a magnetic field of $780 \,\mathrm G$. The atoms have magnetic quantum numbers  $m_F = + 1/2$ ($\,\left| \uparrow \right\rangle\,$) and $m_F = -1/2$ ($\,\left| \downarrow \right\rangle\,$) and correlate to the $F = 1/2$ hyperfine level of the ground state at $0\,\mathrm G$. They are confined in a harmonic 3D cigar-shaped trapping potential which is generated in radial direction mainly by a focused $1070 \,\mathrm{nm}$ dipole trap laser beam and along the axial direction mainly by a magnetic field gradient. The temperature $T$ is set via evaporative cooling and is measured by fitting a distribution obtained from the second order quantum virial expansion to the outer wings of the density profile \cite{LiuThesis}. The particle number $N_{\sigma}$ per spin state ranges from $3\times 10^4$ for the lowest temperature of about $0.3\,T_\mathrm F$ to $3\times 10^5$ for the highest temperature of about $3\,T_\mathrm F$. The  population balance of the two spin states is assured by means of a $100\; \mathrm{ms}$ long resonant RF pulse that mixes the two Zeeman states $\,\left| \uparrow \right\rangle\,$ and $\,\left| \downarrow \right\rangle\,$. For a spin-balanced system the Fermi energy is given by $E_\mathrm F = \hbar (6 N_{\sigma} \omega_r^2\omega_a)^{1/3}$, where $\omega_r$ and $\omega_a$ denote the radial and axial trapping frequency, respectively. In our experiment $\omega_r$ ranges from about $2\pi \times 300\, \mathrm{Hz}$ to $2\pi \times 1.6 \, \mathrm{kHz}$ while $\omega_a = 2\pi \times 21 \, \mathrm{Hz}$ is almost constant as it is dominated by the magnetic confinement. The interaction parameter \kfa can  be tuned by changing either the scattering length $a$ via the broad magnetic Feshbach resonance located at $832 \,\mathrm G$ \cite{zurn2013precise,grimm_feshbach}, or by adjusting the Fermi energy $E_\mathrm F$.

\section{Measuring the pair fraction}
In order to determine the pair fraction $N_p /N_{\sigma}$ we  measure the particle numbers $N_p$ and  $N_{\sigma}$ separately. $N_{\sigma}$ is obtained by means of spin-selective absorption imaging of the $\,\left| \uparrow \right\rangle\,$ component using a $\sigma^-$-polarized $671 \,\mathrm{nm}$ laser beam resonant with the $D_2$ transition of ${}^6 \mathrm{Li}$ \cite{ketterle}. This transition is essentially closed due to a decoupling of the nuclear spin  and  the total electronic angular momentum in the Paschen-Back regime of the hyperfine structure \cite{SelimJochim}. All $\,\left| \uparrow \right\rangle\,$ atoms will be counted regardless whether they are free or bound in the weakly-bound pairs. Since the binding energy $E_b$  of these pairs is always less than $h \times 1\,\mathrm{MHz}$ in our experiments, the imaging laser is resonant with both free atoms and bound pairs. In order to determine the number of bound pairs $N_p$, we transfer all pairs to states that are invisible in our detection scheme and measure again the remaining $\left| \uparrow \right\rangle$ state atom number via absorption imaging. We use two different bound-state transfer methods which produce consistent results. They are briefly described in the following.

\subsection{Optical transfer (OT) method}
\begin{figure}[t]
	\begin{minipage}[t]{0.49\columnwidth}
		\includegraphics[width=\textwidth]{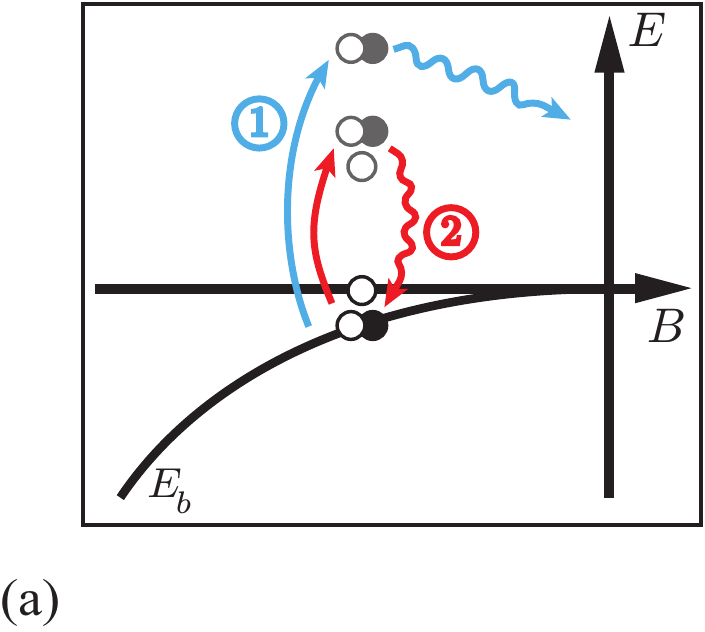}
	\end{minipage}
	\hfill
	\begin{minipage}[t]{0.49\columnwidth}
    	\includegraphics[trim = 0 0 2.6in 0,clip, width=\textwidth]{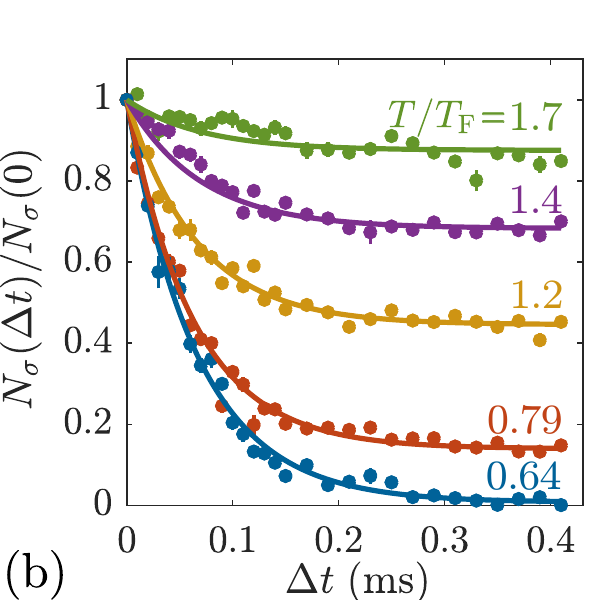}
	\end{minipage}\vspace{0.01cm}
	\begin{minipage}[b]{0.49\columnwidth}
		\includegraphics[width=\textwidth]{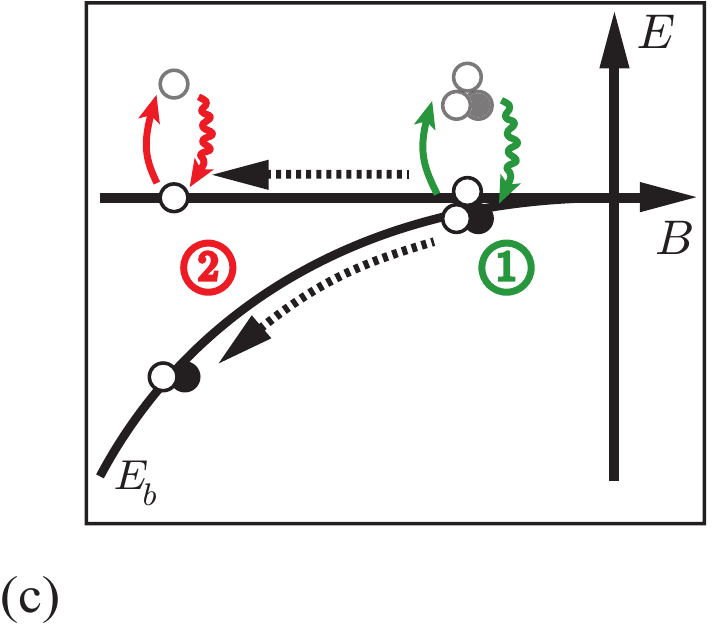}
	\end{minipage}
	\hfill
	\begin{minipage}[b]{0.49\columnwidth}
	\includegraphics[trim = 2.5in 0 0 0 0,clip, width=\textwidth]{figure_2bd.pdf}
	\end{minipage}
	\caption{Measurement of the number of fermion pairs. (a,b) Optical transfer method. A resonant laser pulse transfers pairs to states which are invisible to our detection scheme [blue arrows (1)]. The total number $N_\sigma(\Delta t)$ of remaining fermion pairs and single atoms is measured by absorption imaging [red arrows (2)]. (b) shows  $N_\sigma(\Delta t)/N_\sigma(0)$ as a function of the pulse width $\Delta t$ at a magnetic field of $726 \, \mathrm G$ for various temperatures $T/T_\mathrm F=\left\{0.64,0.79,1.2,1.4,1.7\right\}$. The solid lines are fit curves using Eq.\: \eqref{eq:Nm}. (c,d) Magnetic transfer method. Using absorption imaging, the particle number $N_\sigma = N_a+N_p$ is  measured at the  magnetic field (1) and the number of unbound atoms $N_a$ is measured after a fast ramp to (2). (d) shows the measured particle numbers at (1) ($B =  726 \, \mathrm G$, green solid circles) and at (2) ($B =  550 \, \mathrm G$, red solid squares) for various temperatures $T/T_\mathrm F$.}
	\label{fig:daten}
\end{figure}

This transfer method is based on resonant excitation of fermion pairs to a more strongly bound molecular state ($A^1\!\Sigma_u^+,v^\prime=68$) with a laser ($\lambda = 673 \, \mathrm{nm}$) which is detuned by $2\,\mathrm{nm}$ from the atomic transition, see also \cite{partridge2005molecular}. Subsequently, the excited molecules quickly decay to undetected atomic or molecular states, see Fig\: \ref{fig:daten}(a). This optical excitation of the fermion pairs occurs via an admixture of the molecular bound state $X^1\!\Sigma_g^+,v=38$  to the fermion pair wave function \cite{partridge2005molecular}.

If, for now, we ignore other loss processes, the number of fermion pairs decays exponentially as a function of the laser pulse length $\Delta t$ such that the measured total number $N_\sigma(\Delta t)$ of $m_F = +1/2$ atoms as a function of time is given by
\begin{equation}
	N_\sigma(\Delta t)=N_{\sigma}(0)-N_p \left( 1-\mathrm e^{-k_1 \Delta t} \right),
	\label{eq:Nm}
\end{equation}
where $1/k_1$ is the time constant for the optical excitation. Figure\:\ref{fig:daten}(b) shows this decay for five different initial temperatures $T/T_\mathrm F$  at a magnetic field of $726 \, \mathrm G$. By fitting Eq.\:\eqref{eq:Nm} to the measured data (see fit curves) we are able to extract the pair number $N_p$. Besides the photoexcitation of pairs a loss in $N_\sigma$ could in principle also be induced by photoassociation of two free atoms. However, we made sure that within our field range its rate is negligible. The photoassociation rate constants range between  $1\times 10^{-9} \text{ and } 3\times 10^{-9}\,\mathrm{cm}^5\!\left(\mathrm{W\,s}\right)^{-1}$ for magnetic fields between $726$ and $820\, \mathrm G$. We work with low particle densities of at most $10^{11}\,\mathrm{cm}^{-3}$ and a maximum laser intensity of about $1.9 \,  \mathrm W / \mathrm{cm}^2$.

For the data shown in Fig.\:\ref{fig:daten}(b) the laser intensity is $0.22\, \mathrm{W} / \mathrm{cm}^2$ and the peak density for the lowest temperature of $T/T_\mathrm F = 0.64$ is $1.4 \times 10^{11}\,\mathrm{cm}^{-3}$ which corresponds to an initial photoassociation time constant of about $33\,\mathrm{ms}$. This is much longer than the loss dynamics observed in Fig.\:\ref{fig:daten}(b). Indeed, the fact that the curves in Fig.\:\ref{fig:daten}(b) approach constant values for pulse times $ t \gtrsim 0.3 \, \mathrm{ms}$ already suggests that the photoassociation of free atoms is negligible.

However, closer to resonance the time constants for photoassociation and pair excitation become more comparable. Therefore, we generally release the particles from the trap $0.3 \, \mathrm{ms}$ before applying the laser pulse. The subsequent expansion lowers the cloud density by about a factor of 4 and  assures additionally that photoassociation is negligible. Furthermore,  lowering the density also strongly suppresses regeneration of depleted Fermi pairs during the laser pulse, since pair regeneration mainly occurs via three-body recombination. We have checked that during the expansion the fermion pairs do not break up. For this, we carried out measurements at a magnetic field of $780\, \mathrm{G}$, working at the lowest temperatures of about $0.3\, T_\mathrm F$, where only about $10-15\,\%$ of the atoms are unbound  and thus photoassociation does not play a significant role. We measured the same pair numbers with and without expansion.

In general the OT method works very well up to magnetic fields of about $B = 820\, \mathrm{G}$, close to the Feshbach resonance. There, we observe marked deviations from the exponential decay in Eq. \eqref{eq:Nm}, a behavior, that also had been reported earlier by the Rice group \cite{partridge2005molecular}. An analysis of these signals would require a better understanding of the nature of strongly interacting pairs. For this reason, we decide to stay below magnetic fields of $820 \, \mathrm{G}$ for the present investigations where the analysis is unequivocal.

\subsection{Magnetic transfer (MT) method}
Here, we increase the binding energy of the pairs to $h\times 80.6 \, \mathrm{MHz}$ by quickly ramping the magnetic field  at $20 \, \mathrm{G/ms}$ down to $550 \, \mathrm G$, see Fig.\:\ref{fig:daten}(c). This works very efficiently without breaking up the molecules as previously shown in \cite{Regal2004,Zwierlein2004}. At 550 G the fermion pairs cannot be resonantly excited anymore by the imaging laser and become invisible to our detection scheme, see \cite{grimm}. $N_p$ is determined as the difference of the numbers for atoms and pairs ($N_\sigma$) measured before the ramp and unbound atoms ($N_a$) obtained after the ramp. Figure\:\ref{fig:daten}(d) shows these particle numbers for different temperatures at a magnetic field of $726 \, \mathrm G$.

\begin{figure}[h]
\includegraphics[width=\columnwidth]{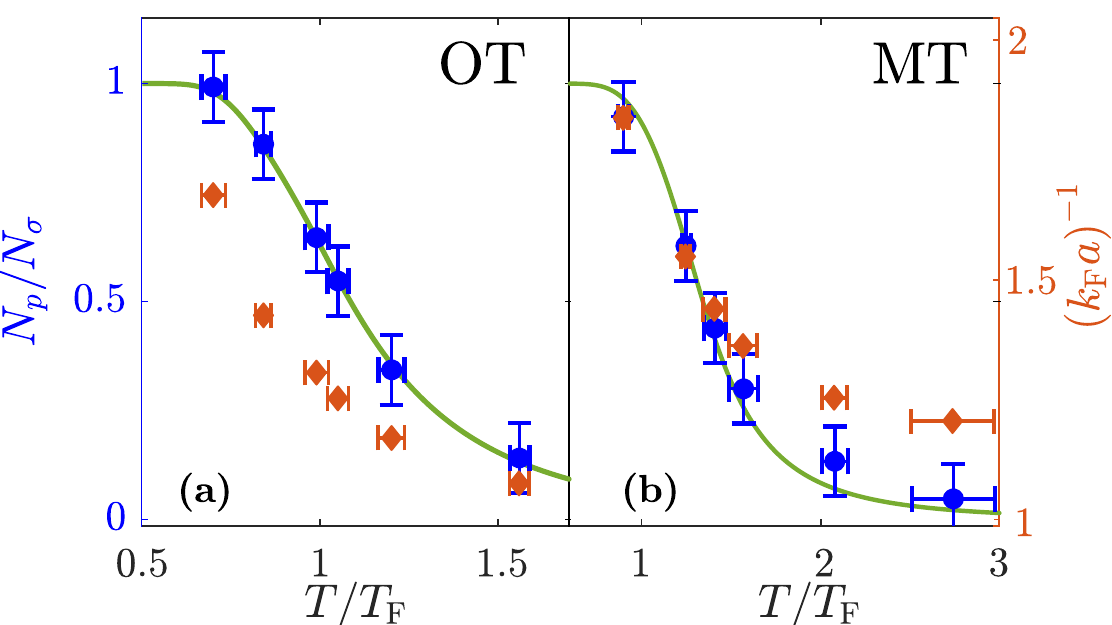}
\caption{Measured pair fractions $N_p/N_\sigma$ (blue circles)  at $726\,\mathrm G$ for various temperatures $T/ T_\mathrm F$. (a) Optical transfer (OT) method, (b) Magnetic transfer (MT) method (see Fig.\,\ref{fig:daten}). We note that due to evaporative cooling \kfa also changes with $T/ T_\mathrm F$ (orange diamonds).  The green curves are calculations based on the classical model.}
\label{fig:auswertung}
\end{figure}

We did not perform measurements with the MT method for magnetic fields higher than $750 \, \mathrm G$ because of technical limitations for the ramping speed. If the field ramp duration  ($\approx 10\,\mathrm{ms}$ for the case of $750\,\mathrm{G}$) becomes comparable to the equilibration time for the atom-molecule mixture  (a few milliseconds at $750\,\mathrm{G}$) the measurement does not yield the correct molecule number anymore. This restriction of the magnetic field ramp implies that we cannot use the MT method in the strong interaction crossover regime, but only in the far BEC regime. There, however, the MT method is quite useful to check for consistency with the OT method. This consistency is shown in Fig.\:\ref{fig:auswertung} where we plot the pair fractions $N_p/N_{\sigma}$ obtained at $726\,\mathrm G$ from both methods as a function of the temperature (blue circles). Since the temperature was adjusted by varying the evaporative cooling, different temperatures correspond to different particle numbers $N_{\sigma}$ and thus to different interaction parameters \kfa (orange diamonds). The green lines are calculated pair fractions using  the classical model. In general, we find good agreement between the experimental data and the theoretical prediction, which also indicates consistency between the OT and MT methods.

\section{Results}

\begin{figure}[t]
\includegraphics[width=\columnwidth]{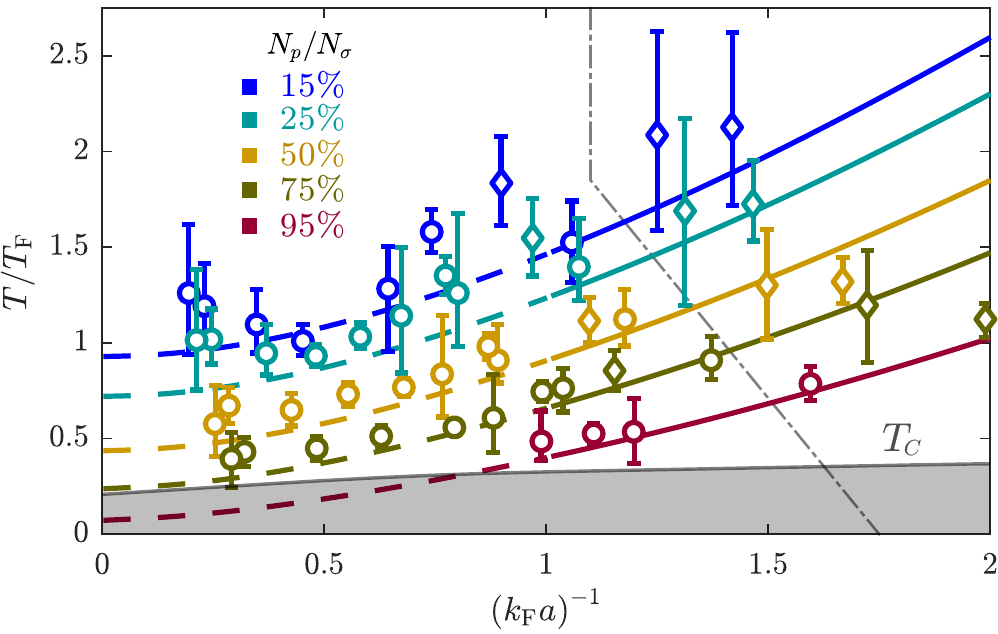}
\caption{Map of the pair fraction $N_p/N_\sigma$ as a function of temperature and interaction strength on the BEC-side of the Feshbach resonance. The circles (diamonds) are measurements obtained with the OT (MT) method. The thick solid and dashed lines are classical model calculations (cf.~Fig.~\:\ref{fig:skizze}). They are dashed in the strong-interaction regime where the classical model is expected to be no longer valid. The error bars include both a statistical and a systematic part, i.e. the standard deviation of the mean of 10 temperature measurements and the uncertainty in determining the molecule fraction from the fit, respectively. The upper-right area bounded by the gray dash-dotted line exhibits $> 5\,\%$  particle loss due to inelastic collisions on the timescale of a measurement. The gray shaded area indicates the superfluid phase below $T_c$, as in Fig.\:\ref{fig:skizze}.}
\label{fig:result}
\end{figure}

We now apply the OT and MT methods to map out  the fraction of pairs on the BEC-side. For this, we perform measurements for a variety of magnetic fields and temperatures. The pair fractions $N_p / N_{\sigma}$ obtained from both experimental methods are shown in Fig.\:\ref{fig:result} (circles: OT method, diamonds: MT method).  The area on the right hand side of Fig.\:\ref{fig:result}, as bounded by the thin dash-dotted line, marks a region where we observe non-negligible loss of particles ($>5\, \%$) during our measurements due to inelastic collisions of bound pairs. This loss increases with \kfa, see e.g.  \cite{salomon_mol-mol, petrov_mol-mol}. In order to simplify our discussion we only consider data points outside this area.

The solid/dashed lines in Fig.\:\ref{fig:result} represent the statistical mechanics model without any adjustable parameters. For higher temperatures we generally observe larger fluctuations and thus larger error bars, because of the larger atom cloud within a limited field-of-view. Overall, we find that \color{black} the agreement between measurement and model remains quite good  even in the crossover regime where this model of classical particles with no interaction energy should be expected to break down. In fact, the model could be expected to work to the extent that the internal degrees of freedoms of the fermion pairs are frozen and only the degrees of freedom associated with the center-of-mass of the pair remain active. This approximately occurs when the fermionic chemical potential changes sign which, using a $t$-matrix approach, we estimate to occur at a coupling value of about $(k_F a)^{-1}=0.5$  at $T_c$. This might explain the good agreement found between the model and the experimental data when $(k_F a)^{-1}\gtrsim 0.5$ as well as with the theoretical calculation based on a self-consistent $t$-matrix approach.

\section{Conclusion}

To conclude, we have systematically mapped out the fermion pair fraction in a strongly interacting Fermi gas as a function of both temperature and coupling strength. Our measurements show how pairing of ultracold fermions changes as we move from the BEC regime into the strong interaction regime. We demonstrate a novel method to measure the pair fractions from the near-BEC limit to the pseudogap regime, which is based on a number measurement of fermion pairs. This method is complementary to existing excitation-gap measurements and has no counterpart in conventional condensed matter systems. We find that a statistical mechanics model treating the fermions and pairs as classical particles  describes the measured data quite well in the investigated range, as we have also confirmed through an advanced many-body calculation based on a $t$-matrix approach. In the future, we plan to extend our measurements and investigate more in detail the coupling region ($0.1\lesssim (k_F a)^{-1} \lesssim 0.5$) where the preformed-pair and the pseudogap regimes overlap with each other.

\acknowledgments

We gratefully acknowledge discussions with Q. Chen and K. Levin and support from the Deutsche Forschungs\-gemeinschaft within SFB/TRR 21 (project part B4), the Baden-W\"{u}rttemberg Foundation, and the Center for Integrated Quantum Science and Technology ($\mathrm{IQ}^{ST}$). C. Chin acknowledges support from the Alexander v. Humboldt fellowship, the Chicago Materials Research Science and Engineering Center by the National Science Foundation (DMR-1420709), and the Army Research Office (W911NF-14-1-0003).

\appendix

\section{Model of a canonical ensemble of non-interacting atoms and molecules} \label{app:model}

In our simple statistical mechanics model we treat the cold gas of fermions and fermion pairs as a classical canonical ensemble of  atoms and molecules, respectively, with negligible interaction energy among each other. In collisions a pair of  $| \uparrow \rangle$ and $| \downarrow \rangle$ atoms can combine to form a molecule, and vice versa a molecule can break up into an unbound pair of $| \uparrow \rangle$, $| \downarrow \rangle$ atoms.
At a given temperature the atom and molecule numbers are in chemical equilibrium. Following \cite{chin}, the equilibrium condition is derived by minimizing the Helmholtz free energy $F=k_\mathrm B T \ln Z$, subject to the constraint of particle number conservation. Here
\begin{equation*}
Z = \frac{{Z_s}^{2 N_a} {Z_s}^{N_p} \mathrm{e}^{N_p E_b / k_\mathrm B T}}{N_a! N_a! N_p!}
\end{equation*}
is the partition function of the system and $Z_s$ and $Z_s e^{-E_b/k_\mathrm B T}$ are the single-particle partition functions for atoms and molecules, respectively. $\overline{\omega} = \sqrt[3]{\omega_r^2\omega_a}$ is the geometric mean of the trapping frequencies $\omega_a, \omega_r$ in axial  and in radial direction, respectively. Using Stirling's formula to approximate the factorials a minimum in the free energy is found at a molecule (pair) number
\begin{equation*}
	N_p = \frac{1}{Z_s}{N_a}^2 \, \mathrm e^{-E_b/k_\mathrm B T} \; ,
\end{equation*}
for a given temperature $T$ and binding energy $E_b = -\hbar^2/(m a^2)$. Using the partition function $Z_s=\left(k_\mathrm B T/\hbar \, \overline{\omega} \right)^3$, the Fermi energy $E_\mathrm F = k_\mathrm B T_\mathrm F = \hbar \overline{\omega} \sqrt[3]{6N_\sigma}$, and the total pair fraction per spin state $N_\sigma = N_a + N_p$ we obtain the following implicit expression for the pair fraction $N_p/N_{\sigma}$ in thermal equilibrium:
\begin{equation*}
\frac{\left(1 - N_p/N_{\sigma} \right)^2}{N_p/N_{\sigma}} = 6 \left( \frac{T}{T_\mathrm F}\right)^3
\exp\left[\frac{E_b}{k_\mathrm B T}\right]\;.
\end{equation*}

\section{Measurements close to unitarity} \label{app:meastounit}

\begin{figure}[htb]
	\includegraphics[width=\columnwidth]{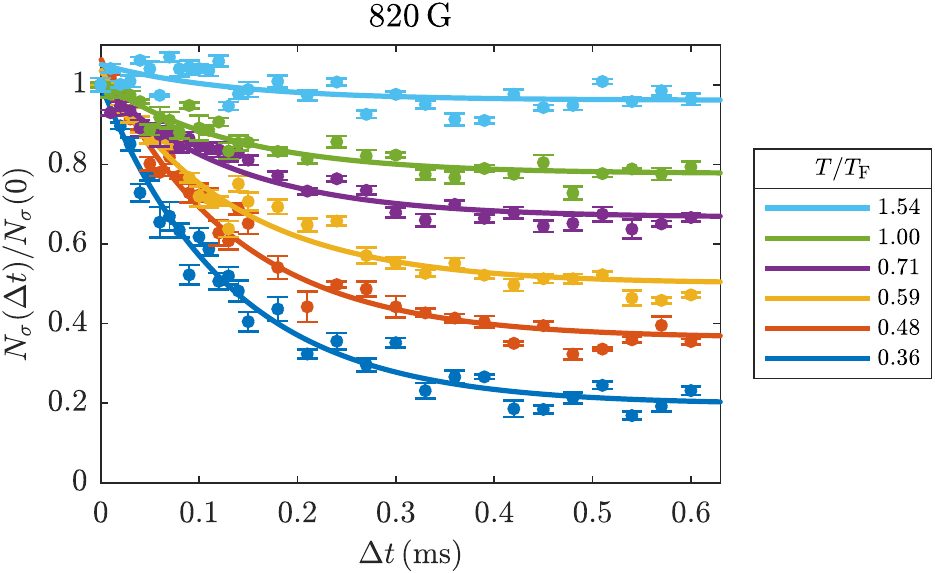}
	\caption{Pair fraction $N_\sigma (\Delta t) / N_\sigma (0)$ after an optical transfer pulse of length $\Delta t$ at a magnetic field of $820 \,\mathrm G$ for various temperatures (see legend). The solid lines are fits of an exponential decay towards a constant offset.}
	\label{fig:820}
\end{figure}

As pointed out in the main text we only carry out measurements at magnetic fields of up to $820\,\mathrm G$ because for higher magnetic fields we observe deviations from an exponential decay during the optical excitation of the pairs towards deeply bound molecules. Such deviations are indeed expected  close to resonance as a result of many body effects \cite{Castin_contact}. In addition, as the optical excitation cross section decreases towards the resonance its rate becomes increasingly comparable to the one of photoassociation. In order to clarify that an exponential fit towards a constant value is still a good description at $820\,\mathrm G$, we show corresponding decay curves in Fig.~\ref{fig:820}. A slight non-exponential behavior of the measured decay will increase the uncertainty in the measured equilibrium pair fraction.

\bibliography{references}

\end{document}